\documentclass[journal]{IEEEtran}

\usepackage{array}
\usepackage{amsmath}
\usepackage{amssymb}
\usepackage{booktabs}
\usepackage{color}
\usepackage{cite}
\usepackage{graphicx}
\usepackage{multirow}
\usepackage{makecell}
\usepackage{subcaption}
\usepackage{tabularx}
\usepackage{url}

\allowdisplaybreaks

\ifCLASSINFOpdf

\else

\fi

\begin{document}

\title{Grid-Forming AI Data Centers: Coordinated Grid Support Control from IT Loads and Energy Storage}

\author{
\IEEEauthorblockN{
Xuechun Zhang\textsuperscript{1,2},
Zhiyi Zhao\textsuperscript{2},
Fei Teng\textsuperscript{2},
Zhen Wang\textsuperscript{1,*}
} \\
\IEEEauthorblockA{
\textsuperscript{1}\textit{College of Electrical Engineering, Zhejiang University, Hangzhou, China} \\
\textsuperscript{2}\textit{Department of Electrical and Electronic Engineering, Imperial College London, London, UK} \\
Emails: xuechun\_zhang@zju.edu.cn, z.zhao26@imperial.ac.uk, f.teng@imperial.ac.uk, eezwang@ieee.org.
}

        % <-this % stops a space
\thanks{*Corresponding author: Zhen Wang (Email: eezwang@ieee.org)

This work was supported by the Smart Grid National Science and Technology Major Project (2025ZD0804300).
}}

\markboth{}%
{Shell \MakeLowercase{\textit{et al.}}: Bare Demo of IEEEtran.cls for IEEE Journals}

\maketitle

\begin{abstract}
To exploit the grid-support capability of AI data centers~(AIDCs), this paper proposes a grid-forming~(GFM) control scheme which coordinates the IT loads and energy storage~(ES) in the AIDC. First, the framework of a grid-connected AIDC is presented, with the multi-stage power conversion from the grid to the IT loads presented. A dual GFM scheme is then proposed for the back-to-back converters to provide grid support through coordinated power adjustments from IT loads and ES. In addition, an aggregated power response model is established to characterize the IT-load flexibility considering multiple IT servers executing diverse tasks. Time-domain simulations of a 100~MW grid-connected AIDC demonstrate that the proposed control scheme can effectively support the grid.
\end{abstract}


\IEEEpeerreviewmaketitle

\section{Introduction}
\IEEEPARstart{A}{rtificial} intelligence data centers (AIDCs) are emerging as large electrical loads in modern power systems. With AIDCs reaching gigawatt-level capacity, the global electricity consumption of data centers is expected to reach 950~TWh by 2030~\cite{Chap1_IEA_2026_EnergyAI}. For AIDCs with converter-based grid interfaces, their dynamic behaviors are shaped by converter control, enabling them to provide ancillary services such as frequency regulation~\cite{Chap1_FFR2_2025,Chap1_FFR_2021,Chap1_PFR_Pcap_2026, Chap1_SFR_2026}. 

%The key is how to exploit the flexibility of IT-load to realize active grid support

Existing studies have explored the frequency support provided by AIDCs, including fast~\cite{Chap1_FFR2_2025,Chap1_FFR_2021}, primary~\cite{Chap1_PFR_Pcap_2026}, and secondary~\cite{Chap1_SFR_2026} frequency response. A droop-based fast frequency response (FFR) strategy is proposed in \cite{Chap1_FFR2_2025}, where IT servers and uninterruptible power supply (UPS) batteries respond to the measured local frequency deviations, while a model predictive control-based FFR is utilized in~\cite{Chap1_FFR_2021}. In~\cite{Chap1_PFR_Pcap_2026}, a droop curve with a deadband is designed to generate GPU power caps, and hardware experiments demonstrate that the power response is sufficiently fast for primary frequency support. An LLM inference task scheduling strategy is proposed in \cite{Chap1_SFR_2026} to track AGC signals in the ancillary service market. However, under grid-following (GFL) interfaces, the frequency support strategies largely rely on phase-locked loop (PLL)-based frequency measurement and feedback-based active power regulation, limiting the speed and effectiveness of frequency support.

Compared with GFL grid interface, grid-forming (GFM) control enables the grid-side converter (GSC) to exhibit voltage-source characteristics \cite{Chap1_GFL_GFM_duality}, providing inherent frequency support without relying on frequency measurement or external power regulation signal. In~\cite{Chap1_UPS_3Mode_GFM_2025}, GFM control is applied to the dedicated GSC of the centralized UPS, enabling the UPS to operate in DC-bus regulation, fault ride-through, and frequency support modes. Nevertheless, how to enable IT loads to operate in GFM mode and coordinate with the UPS to support the grid still remains to be investigated.
%this paper utilizes GFM control for GSC to provide an inherent frequency and voltage response, while coordinating load-side response and ES operation to maintain internal power balance.

To operate under GFM control, the load should meet three requirements~\cite{Chap1_GFM_HE_2023}: 

1) converter-based grid interfaces; 

2) sufficient flexibility for power regulation; 

3) sub-second response to grid changes.

For AIDCs, first, when back-to-back converters serve as grid interfaces of the IT loads, GFM control can be applied to GSC to support the grid~\cite{Chap1_Review_3Modes_2026}. Second, the load flexibility can be provided through power capping~\cite{Chap1_PFR_Pcap_2026}, task scheduling~\cite{Chap1_SFR_2026}, server activation~\cite{Chap1_WakeSleep_2019}  and load migration~\cite{Chap1_Spatial_EmeraldAI_Arxiv} as summarized in Table~\ref{tab:server_control_methods}. Among these methods, power capping enables rapid power regulation by adjusting the GPU power limits dynamically. Through dynamic voltage and frequency scaling (DVFS), the server power can track the power cap on a sub-second timescale~\cite{Chap1_DVFS_control}. These characteristics make AIDCs suitable for GFM control.

\begin{table}[b]
\centering
\caption{Power regulation methods for IT loads.}
\label{tab:server_control_methods}
\footnotesize
\setlength{\tabcolsep}{2pt}
\begin{tabularx}{\columnwidth}{@{}
>{\centering\arraybackslash}m{2.5cm}
>{\centering\arraybackslash}m{4.5cm}
>{\centering\arraybackslash}X
@{}}
\toprule
Method & Mechanism & Timescale \\
\midrule
Power Capping
& Limits the server power through hardware adjustments (DVFS).
& ms--s\textsuperscript{\cite{Chap1_PFR_Pcap_2026}} \\

Task Scheduling
& Regulates power by deferring or pausing delay-tolerant tasks.
& s\textsuperscript{\cite{Chap1_SFR_2026}} \\

Server Activation
& Adjusts the number of active servers by switching on/off states.
& s--min\textsuperscript{\cite{Chap1_WakeSleep_2019}} \\

Load Migration
& Shifts workloads to other AI data centers with available power.
& min--h\textsuperscript{\cite{Chap1_Spatial_EmeraldAI_Arxiv}} \\

\bottomrule

\end{tabularx}
\end{table}

%dynamic voltage and frequency scaling (DVFS)-based power capping~\cite{Server_Power_Capping_2012} and graphics processing unit (GPU) core allocation~\cite{GPU_Grid_Regulation_2026}, enabling data centers to adjust their power consumption for grid regulation. In recent studies, this flexibility has been used for regulation reserve provision via server power modulation~\cite{EnergyQARE_2019}, and frequency regulation via coordinated workload scheduling and uninterruptible power supply (UPS) operation~\cite{Cloud_Frequency_Support_2020}.

% this paper explores the feasibility of implementing the GFM control in AIDCs
On this basis, a GFM control scheme for AIDC is proposed in this paper which employs dual GFM control for the back-to-back converters and coordinates the IT loads and UPS energy storage (ES) to support the grid.

The remainder of the paper is organized as follows. Section II introduces the framework of AIDC, Section III proposes an AIDC GFM control scheme that coordinates IT loads and ES, Section IV characterizes the aggregated power response of IT loads, and Section V presents the simulation results.

\section{AIDC Framework}
\label{sec:AIDC_framework}

\begin{figure}[!t]
\centering
\includegraphics[width=1\linewidth]{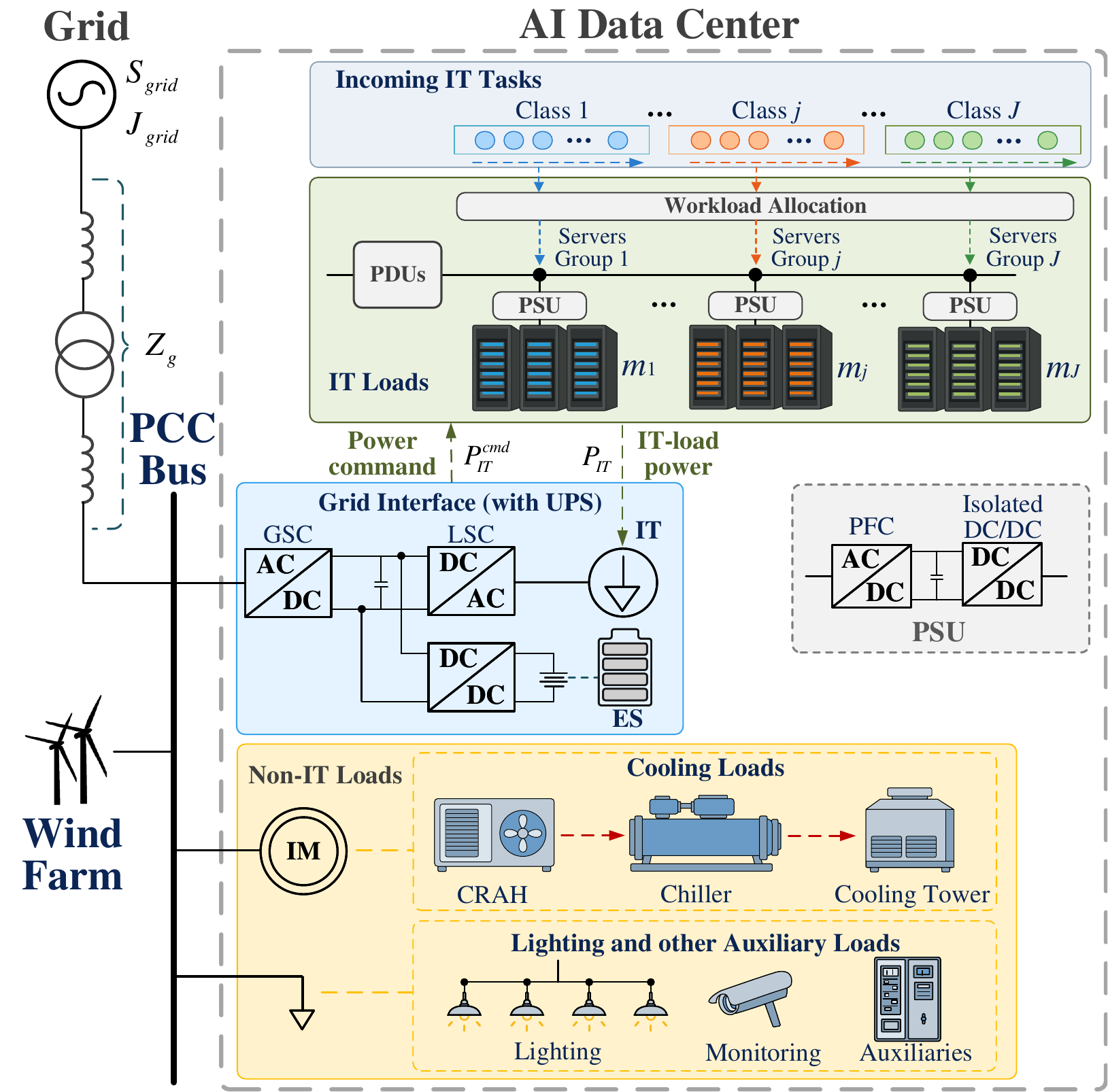}
\caption{Framework of a grid-connected AI data center.}
\label{fig:structure}
\end{figure}

In AIDCs, the UPS can operate in double-conversion mode, bypass mode and battery-backup mode according to~\cite{Chap1_Review_3Modes_2026}. Among them, the double-conversion mode is adopted in this work to implement GFM control on the back-to-back converters. Fig.~\ref{fig:structure} shows the framework of a grid-connected AIDC from an electrical perspective.  

As shown in Fig.~\ref{fig:structure}, IT servers executing different tasks can be modeled as a controllable load, with its aggregated power response model presented in Section~\ref{sec:IT_load_response}. In double-conversion mode, power from the point of common coupling (PCC) is supplied to the IT servers through multi-stage conversion:

1) At the grid interface, the AC/DC GSC and DC/AC load-side converter~(LSC) are connected through a DC link. The ES of the UPS is also connected to the DC link through a DC/DC converter. The proposed control scheme is detailed in Section~\ref{sec:AIDC_control}. 

2) At the server racks, the power distribution units~(PDUs) distribute AC power to the power supply units~(PSUs) of individual servers. Each PSU consists of an AC/DC converter with power-factor control~(PFC) and an isolated DC/DC converter to supply the regulated DC power to the IT loads.

AIDCs also include non-IT loads such as cooling and auxiliary loads, which are modeled as motors and constant-power loads, respectively. The total power consumption is
\begin{equation}
P_{\mathrm{AIDC}}
=
P_{\mathrm{IT}} 
+
P_{\mathrm{ES}} 
+
P_{\mathrm{cool}}
+
P_{\mathrm{aux}},
\label{eq:AIDC_incremental_power}
\end{equation}
where $P_{\mathrm{AIDC}}$ denotes the total AIDC power consumption, $P_{\mathrm{IT}}$, $P_{\mathrm{cool}}$, and $P_{\mathrm{aux}}$ denote the power consumption of IT, cooling and auxiliary loads respectively, $P_{\mathrm{ES}}$ denotes the charging power of UPS ES.

Since the non-IT loads are either dominated by motors with relatively slow dynamics or non-adjustable, their power consumption is assumed constant when focusing on fast grid support~\cite{Chap1_FFR2_2025}. Therefore, on the fast timescale of interest, the AIDC power response is dominated by the IT loads and the ES power of the UPS
\begin{equation}
\Delta P_{\mathrm{AIDC}}
\approx
\Delta P_{\mathrm{IT}} 
+
\Delta P_{\mathrm{ES}},
\label{eq:AIDC_incremental_power}
\end{equation}
where $\Delta P_{\mathrm{IT}}$ and $\Delta P_{\mathrm{ES}}$ denote the power adjustments from IT loads and UPS ES, respectively.

\section{AIDC GFM Control Scheme}
\label{sec:AIDC_control}

In this section, an AIDC GFM control scheme is proposed which exploits both the IT-load flexibility and the UPS energy storage to jointly support the grid.

\begin{figure}[!t]
  \centering
  \includegraphics[width=1\linewidth]{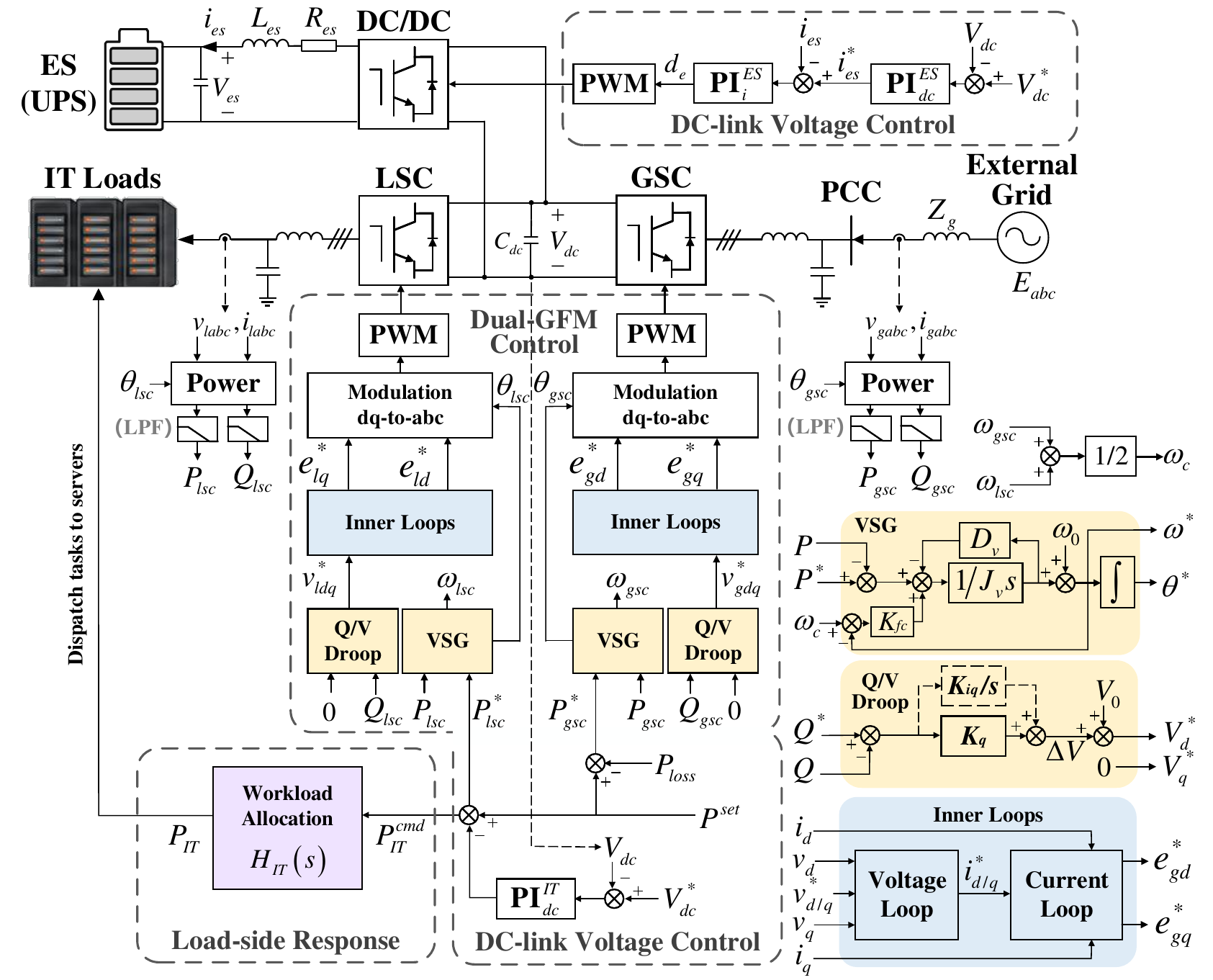}
  \caption{Diagram of the proposed GFM control scheme.}
  \label{fig:control_AIDC}
\end{figure}

Fig.~\ref{fig:control_AIDC} shows the dual GFM control scheme, in which GFM control is applied to both the GSC and LSC. Through dc-link voltage control, a coordinated control strategy is further implemented to jointly adjust the IT-load power consumption and UPS ES power. In this work, power is defined as positive when it is consumed by loads or absorbed by ES.

\subsection{Dual GFM Control for Back-to-Back Converters}
\label{sec:GFM}

In the dual GFM control scheme, the GSC and LSC employ the same GFM control structure, i.e., a virtual synchronous generator (VSG)-based active-power loop, a reactive-power/voltage (Q/V) droop loop, and cascaded voltage and current inner loops. Details of the GFM control can be found in~\cite{Chap3_GFM_control}.

On the GSC side, the converter under VSG-GFM control behaves as a voltage source that inherently supports the grid frequency by providing virtual inertia and damping. Its active power response to grid-side disturbance $\Delta\omega_g(s)$ is given by
\begin{equation}
G_{p\omega}(s)
=
\frac{\Delta P_{gsc}(s)}{\Delta\omega_g(s)}
=
(J_{v}s+D_{v})(1+\tau_{p}s),
\label{eq:GSC_Pf_response}
\end{equation}
where \(J_{v}\) and \(D_{v}\) are the virtual inertia and damping coefficients of GSC, and $\tau_{p}$ denotes the time constant of the low-pass filter for active-power measurement.

On the LSC side, the GFM converter provides stable voltage and frequency for IT loads. Its active-power reference is generated by the dc-link voltage PI controller
\begin{equation}
P_{lsc}^\star(s)
=
P^{set}/s
+
G_{dc}^{IT}(s) \cdot
\left[V_{dc}(s)-V^\star_{dc}/s \right],
\label{eq:P_lsc_ref_s}
\end{equation}
where $V_{dc}(s)$ and $V^\star_{dc}$ denote the dc-link voltage and its reference value, $P^{set}$ denotes the load power setpoint, and $G_{dc}^{IT}(s)=K_{p,dc}^{IT}+K_{i,dc}^{IT}/s$ is the dc-voltage PI controller on IT-load side.

\subsection{Power Response of IT loads}
\label{sec:load_response}
The power command for IT loads also follows the power reference given in \eqref{eq:P_lsc_ref_s},
i.e., \(P_{IT}^{cmd}(s)=P^\star_{lsc}(s)\). The IT-load power response can be approximated as a first-order lag model (see Section~\ref{sec:IT_load_response})
\begin{equation}
H_{IT}(s)
=
\frac{\Delta P_{IT}(s)}
{\Delta P_{IT}^{cmd}(s)}
\approx
\frac{1}{1+s\tau_{\mathrm{eq}}},
\label{eq:IT_eq_response}
\end{equation}
where $\tau_{eq}$ denotes the equivalent time constant of IT loads. 

Neglecting the converter switching losses, the dc-link voltage dynamics based on the power balance among the GSC, LSC, and the DC/DC converter of ES can be linearized as
\begin{equation}
 \Delta V_{dc}(s)
 =G_{vp}(s) \cdot \big[\Delta P_{gsc}(s)-\Delta P_{lsc}(s)-\Delta P_{es}(s)\big],
 \label{eq:Vdc_dynamics}
\end{equation}
where $G_{vp}(s)=1/(sC_{\mathrm{dc}}V_{dc}^\star)$.

Combining \eqref{eq:GSC_Pf_response}--\eqref{eq:Vdc_dynamics}, the IT-load response $\Delta P_{IT}(s)$ to grid disturbance $\Delta\omega_g(s)$ without ES participation is derived as
\begin{equation}
\frac{\Delta P_{IT}(s)}{\Delta\omega_g(s)}
=
G_{p\omega}(s)
\frac{G_{vp}(s)G_{dc}^{IT}(s)H_{IT}(s)}
{1+G_{vp}(s)G_{dc}^{IT}(s)H_{IT}(s)}.
\label{eq:tf_IT_power_response}
\end{equation}

\subsection{Coordinated Power Response of IT Loads and UPS ES}
\label{sec:coordination}

To fully exploit the grid-support capability of AIDC, the power of UPS ES is adjusted in coordination with IT-load power regulation through dc-link voltage control.

Similarly, the current reference of the ES in UPS is generated by a dc-link voltage PI controller
\begin{equation}
i_{es}^\star(s)
=
G_{dc}^{ES}(s) \cdot
\left[V_{dc}(s)-V^\star _{dc}\right / s] / V_{es},
\label{eq:ES_dc_voltage_PI}
\end{equation}
where $i_{es}^\star(s)$ denotes the ES current reference, $G_{dc}^{ES}(s)=K_{p,dc}^{ES}+K_{i,dc}^{ES}/s$ denotes the ES dc-voltage PI controller, $V_{es}$ denotes the ES voltage, which is considered constant over the normal operating SOC range.

\begin{figure}[!t]
  \centering
  \includegraphics[width=1\linewidth]{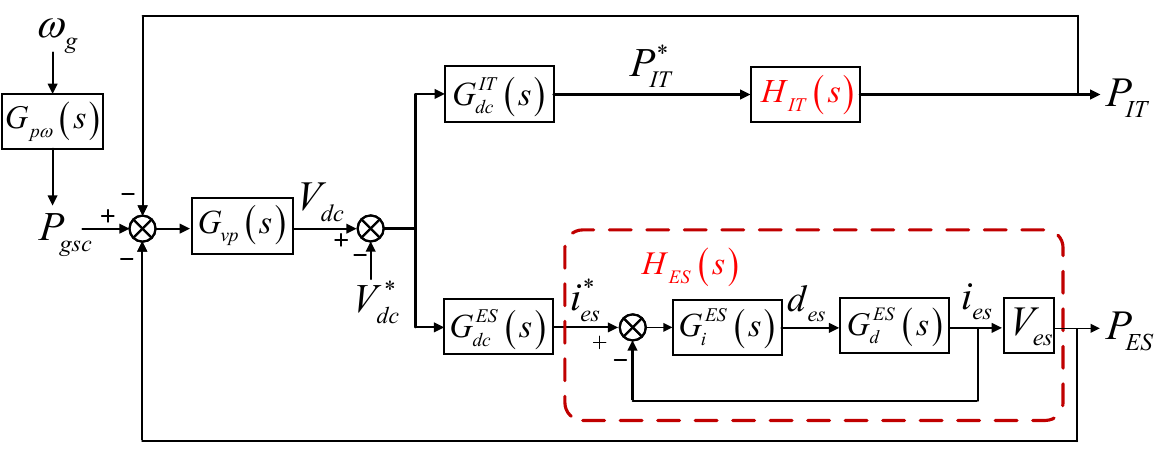}
  \caption{Coordinated control of IT loads and ES.}
  \label{fig:Pdc_transfer_function}
\end{figure}

As shown in Fig.~\ref{fig:Pdc_transfer_function}, with coordinated control of IT loads and ES, the closed-loop transfer functions from the grid disturbance $\Delta\omega_g(s)$ to their power responses are derived as
\begin{equation}
\frac{\Delta P_{IT}(s)}{\Delta\omega_g(s)}
=
\frac{
G_{p\omega}G_{vp}\cdot G_{dc}^{IT}H_{IT}
}{
1+G_{vp}\left(
G_{dc}^{IT}H_{IT}+G_{dc}^{ES}H_{ES}
\right)
},
\label{eq:P_IT_response}
\end{equation}

\begin{equation}
\frac{\Delta P_{ES}(s)}{\Delta\omega_g(s)}
=
\frac{
G_{p\omega}G_{vp}\cdot G_{dc}^{ES}H_{ES}
}{
1+G_{vp}\left(
G_{dc}^{IT}H_{IT}+G_{dc}^{ES}H_{ES}
\right)
},
\label{eq:P_ES_response}
\end{equation}
where the argument $(s)$ is omitted for brevity and
\begin{equation*}
\begin{aligned}
H_{ES}(s)
&=
\frac{G_i^{ES}G_{d}^{ES}}
{1+G_i^{ES}G_{d}^{ES}},\\
G_i^{ES}(s)
&= (K_{p,i}^{es}+K_{i,i}^{es}/s)/V_{dc}^{\star},\\
G_{d}^{ES}(s)
&= V_{dc}^{\star}/\left(R_{es}+sL_{es}\right).
\end{aligned}
\end{equation*}

To quantify the contribution of the IT loads to the total power response, the power-sharing ratio \(\rho(s)\) is defined as
\begin{equation}
\rho(s)
\triangleq
\frac{
\Delta P_{IT}(s)
}{
\Delta P_{IT}(s)+\Delta P_{ES}(s)
}
\in[0,1].
\label{eq:rho}
\end{equation}

Based on \eqref{eq:P_IT_response} and \eqref{eq:P_ES_response}, since the ES inner loop response \(H_{ES}\) and the IT-load response \(H_{IT}\) are much faster than those of the dc-link voltage loops \(G^{IT}_{dc},G^{ES}_{dc}\)~\cite{Chap3_AIDC_modeling,Chap1_PFR_Pcap_2026}, the power-sharing ratio $\rho$ in \eqref{eq:rho} can be approximated by
\begin{equation}
\rho(s)
\approx
\frac{
K_{p,dc}^{IT}s+K_{i,dc}^{IT}
}{
(K_{p,dc}^{IT}+K_{p,dc}^{ES})s
+
(K_{i,dc}^{IT}+K_{i,dc}^{ES})
}.
\label{eq:rho_approx}
\end{equation}

For a prescribed $\rho_0$, the desired power sharing between IT loads and ES is achieved when control parameters satisfy
\begin{equation}
K_{p,\mathrm{dc}}^{\mathrm{IT}}=
\frac{\rho_0}{1-\rho_0}\,
K_{p,\mathrm{dc}}^{\mathrm{ES}},~
K_{i,\mathrm{dc}}^{\mathrm{IT}}
=
\frac{\rho_0}{1-\rho_0}\,
K_{i,\mathrm{dc}}^{\mathrm{ES}}.
\label{eq:gain_partition}
\end{equation}

\section{IT-Load Power Response Model}
\label{sec:IT_load_response}

To characterize the IT-load response to the power command given in \eqref{eq:P_lsc_ref_s}, in this section, an aggregated IT-load power response model is developed considering multiple IT servers executing diverse tasks.

\subsection{Single-server Power Response}
\label{sec:single_server_response}
The power consumption of an IT server can be described by a function of its utilization and core frequency~\cite{Chap4_P_uf_server_2026}
\begin{equation}
P_s
=
P_s(u,f_s),
\label{eq:server_P}
\end{equation}
where $u\in[0,1]$ denotes the normalized server utilization, and $f_s$ denotes the CPU/GPU core clock frequency.

Among the power regulation methods summarized in Table~\ref{tab:server_control_methods}, the power capping method enables fast power tracking at ramp rates exceeding 1~p.u./s\cite{Chap1_PFR_Pcap_2026}, allowing servers to provide load-side response for grid support control described in Subsection~\ref{sec:load_response}.

The dynamics of the server power regulation can be approximated as a first-order model~\cite{Chap1_FFR_2021}
\begin{equation}
\tau \frac{dP_s}{dt}
=
P_s^{cmd}-P_s, \quad
(1-\alpha) P_s^{max} 
\leq
P_s^{cmd}
\leq 
P_s^{max},
\label{eq:server_power_dynamics}
\end{equation}
where $\tau$ denotes the time constant of server power regulation, $\alpha$ denotes maximum power reduction ratio, $P_s^{\mathrm{cmd}}$ denotes the server power command, and $P_s^{\mathrm{max}}$ is the server power at the highest-efficiency operating point. 

\subsection{Aggregated Power Response Model}
\label{sec:multi_server_response}

Consider an AIDC with $M$ homogeneous servers executing $J$ classes of tasks. The parameters $\tau_j$ and $\alpha_j$ in \eqref{eq:server_power_dynamics} vary across classes, depending on the flexibility of class-$j$ tasks.

Based on \eqref{eq:server_P}, the aggregated IT-load power is
\begin{equation}
P_{\mathrm{IT}}
=
\sum_{j=1}^{J}
m_jP_{s,j}
(u_j,f_{s,j}),
\label{eq:aggregated_IT_power}
\end{equation}
where $m_j$ denotes the number of servers executing tasks in class $j$, $P_{IT} \in [P^{\mathrm{min}}_{IT}, P^{\mathrm{max}}_{IT}]$ is the aggregated IT-load power,
\begin{equation}
P_{IT}^{\min}
=
\sum_{j=1}^{J}m_j(1-\alpha_j)P_s^{\max},
\quad
P_{IT}^{\max}
=
\sum_{j=1}^{J}m_jP_s^{\max}.
\label{eq:Agg_power_range}
\end{equation}

Based on the single-server power response in \eqref{eq:server_power_dynamics}, the aggregated IT-load power response can be approximated as
\begin{equation}
H_{IT}(s)
= \frac
{\Delta P_{IT}(s)}
{\Delta P_{IT}^{cmd}(s)}
=
\sum_{j=1}^{J}
\frac
{m_j\beta_j}
{1+s\tau_j}
\approx
\frac{1}
{1+s\tau_{\mathrm{eq}}},
\label{eq:Agg_power_response}
\end{equation}
where $\beta_j$ denotes the share of change in power command for a class-$j$ server.

If $\Delta P_{IT}^{cmd}$ is allocated among servers in proportion to their available power reduction ratios, the power adjustment for each server in class $j$ is given by
\begin{equation}
\Delta P_{s,j}^{cmd}=\beta_j\Delta P_{IT}^{cmd},
\qquad
\beta_j = \alpha_j /
(\displaystyle\sum_{k=1}^{J}m_k\alpha_k).
\label{eq:Pcmd_allocation fraction}
\end{equation}

Based on \eqref{eq:Pcmd_allocation fraction}, the equivalent time constant $\tau_{eq} $ of the aggregated power response model in \eqref{eq:Agg_power_response} can be derived as
\begin{equation}
\tau_{\mathrm{eq}}
= \sum_{j=1}^{J}m_j\beta_j\tau_j
=
({\sum_{j=1}^{J}m_j\alpha_j \tau_j})
/
({\sum_{k=1}^{J}m_k\alpha_k}).
\label{eq:tau_eq}
\end{equation}

\section{Case Studies}
\label{sec:case_studies}

In this section, case studies are conducted on a grid-connected AIDC test system to validate the proposed GFM control scheme.

\subsection{System Settings}

The AIDC test system is established in MATLAB/Simulink based on the report by Pacific Northwest National Laboratory (PNNL)~\cite{Chap5_PNNL_2026}. As shown in Fig.~\ref{fig:structure}, a 100~MW AI data center and a 100~MW wind farm are connected to a 200~MVA low-inertia grid, with main parameters given in Table~\ref{tab:AIDC_parameters}. 

\begin{table}[!t]
  \centering
  \caption{Main parameters of the AIDC test system~\cite{Chap5_PNNL_2026}.}
  \label{tab:AIDC_parameters}
  \small
  \setlength{\tabcolsep}{3pt}
  \renewcommand{\arraystretch}{1.2}
\begin{tabularx}{\columnwidth}{*{2}{>{\centering\arraybackslash}X}|*{2}{>{\centering\arraybackslash}X}}
  \hline
  Parameter & Value & Parameter & Value \\
  \hline
  $V_{\mathrm{N}}$ & $480\,\mathrm{V}$
    & $f_N$ & $60\,\mathrm{Hz}$ \\
    
  $P_{\mathrm{AIDC}}$ & $100\,\mathrm{MW}$
    & $J_{grid}$ & $3\,\mathrm{s}$ \\

  $V^\star _{dc}$ & $800\,\mathrm{V}$
    & $C_{{dc}}$ & $11.55\,\mathrm{F}$ \\

  $R_g$ & $0.05\,\mathrm{p.u.}$
    & $L_g$ & $0.25\,\mathrm{p.u.}$ \\
       
  $P^{set}$ & $70\,\mathrm{MW}$
    & $S_{ES}$ & $84\,\mathrm{MVA}$ \\

  $P_{\mathrm{cool}}$ & $20\,\mathrm{MW}$
    & $P_{\mathrm{aux}}$ & $10\,\mathrm{MW}$ \\

  $J_v$ & $2\,\mathrm{s}$
    & $D_v$ & $50\,\mathrm{p.u.}$ \\

  $\tau_{p}$ & $20\,\mathrm{ms}$
    & $V_{\mathrm{ES}}$ & $480\,\mathrm{V}$\\

  $K^{IT}_{p,dc}$ & $160\,\mathrm{kW/V}$
    & $K^{IT}_{i,dc}$ & $100\,\mathrm{kW/(V \cdot s)}$\\

  $K^{ES}_{p,dc}$ & $160\,\mathrm{kW/V}$
    & $K^{ES}_{i,dc}$ & $100\,\mathrm{kW/(V \cdot s)}$\\

  $K^{ES}_{p,i}$ & $6.25\,\mathrm{V/kA}$
    & $K^{ES}_{i,i}$ & $~~50\,\mathrm{V/(kA \cdot s)}$ \\
  \hline
\end{tabularx}
\end{table}

In the test system, the AIDC adopts the proposed control scheme as described in Section~\ref{sec:AIDC_control} with a power-sharing ratio of 50\%, while the wind farm operates under GFL control. The IT loads account for 70\% of the total AIDC power, while the ES capacity of UPS is set to 1.2~times the IT-load power~\cite{Chap5_PNNL_2026}. 

According to the report by the Electric Power Research Institute~(EPRI) ~\cite{Chap5_EPRI_2024}, training, fine-tuning and inference tasks account for approximately 30\%, 10\%, and 60\% of the IT-load power consumption as given in Table~\ref{tab:IT_task_classification}. The available power reductions of different tasks are assigned based on their flexibility levels as given in \cite{Chap4_NE_2026}, and the time constants are determined based on a maximum power ramp rate of 1~p.u./s~\cite{Chap1_PFR_Pcap_2026}. Based on \eqref{eq:Agg_power_range} and
\eqref{eq:tau_eq}, the aggregated power response can be described by \eqref{eq:Agg_power_response} within $P_{\mathrm{IT}}\in [0.8P^{set},P^{set}]$ with an equivalent time
constant of $\tau_{\mathrm{eq}} \approx 0.2~\mathrm{s}$.

\begin{table}[!t]
\caption{Classification and Parameters of IT tasks~\cite{Chap5_EPRI_2024,Chap4_NE_2026}.}
\label{tab:IT_task_classification}
\centering
\footnotesize
\setlength{\tabcolsep}{2pt}
\renewcommand{\arraystretch}{1.15}
\begin{tabular*}{\columnwidth}{@{\extracolsep{\fill}}cccc@{}}
\hline
\rule{0pt}{2.6ex}
Task class & Training & Fine-tuning & Inference \\
           &          &             & (Deferrable\textsuperscript{*}) \\[2pt]
\hline
\rule{0pt}{2.6ex}
Share of IT-load power
    & 30\% & 10\% & $60\%\cdot0.5$ \\[2pt]
Available power reduction $\alpha$
    & $50\%$
    & $25\%$
    & $10\%$ \\[2pt]
Response time constant $\tau$
    & $250\,\mathrm{ms}$
    & $125\,\mathrm{ms}$
    & $50\,\mathrm{ms}$ \\[3pt]
\hline
\end{tabular*}

\smallskip
\begin{minipage}{\columnwidth}
\footnotesize
\textsuperscript{*} Assuming deferrable tasks account for 50\% of inference power consumption.
\end{minipage}
\end{table}

Test Scenario: Initially, the test system operates in steady state with power balance. At t = 2~s, a sustained power deficit of $\Delta P_{grid} = 20~\mathrm{MW}$ occurs in the external grid.

\begin{figure}[!b]
  \centering
  \includegraphics[width=\linewidth]{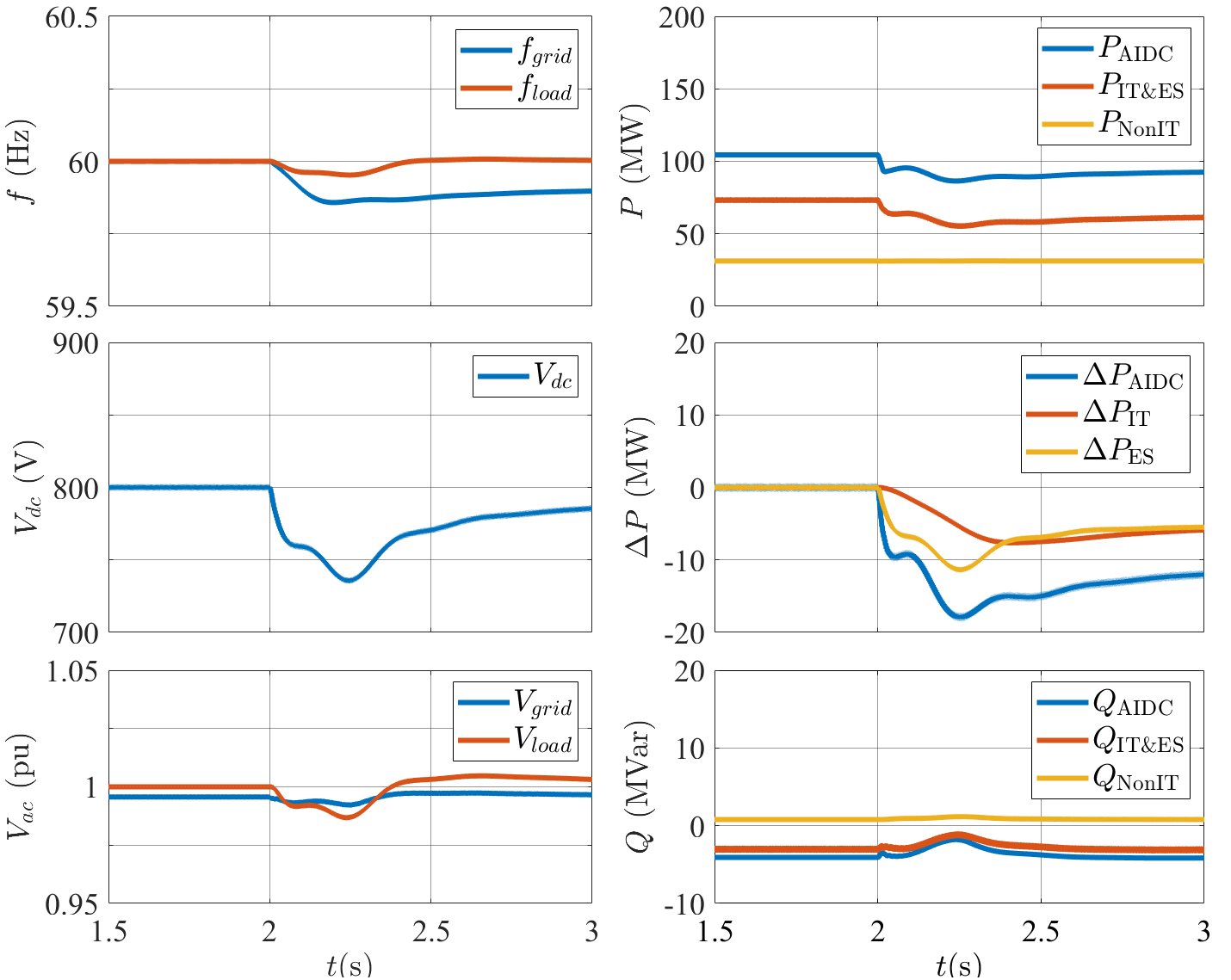}
  \caption{Simulation results of the proposed GFM-AIDC.}
  \label{fig:Scope_AIDC}
\end{figure} 

\subsection{Simulation Results}

As shown in Fig.~\ref{fig:Scope_AIDC}, under the proposed dual GFM control scheme, the AIDC power consumption is dynamically reduced by the GFM-controlled GSC to support the grid frequency during the disturbance, while the load-side stability is supported by the GFM-controlled LSC. Based on the dc-link voltage deviation, the power of IT loads and UPS ES is jointly adjusted. At the beginning, the ES contributes more since the ES power response $H_{ES}(s)$ in~\eqref{eq:P_ES_response}  is faster than the IT-load response $H_{IT}(s)$ in~\eqref{eq:P_IT_response}, while their power-sharing ratio eventually approaches the prescribed $\rho_0 = 50\%$.

\begin{figure}[!t]
  \centering
  \includegraphics[width=\linewidth]{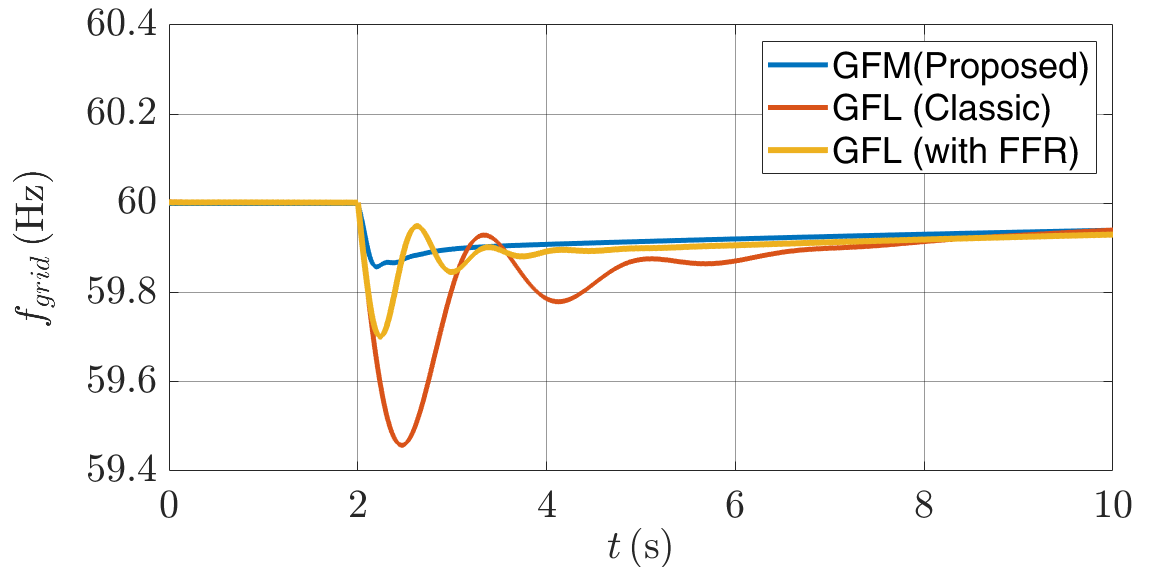}
  \caption{Frequency responses under different control schemes.}
  \label{fig:frequency_response_curves}
\end{figure}

The proposed control scheme is further compared with a) classic GFL control~\cite{Chap5_PNNL_2026} and b) GFL control with FFR~\cite{Chap1_FFR2_2025} over a 10~s time window. As shown in Fig.~\ref{fig:frequency_response_curves} and Table~\ref{tab:frequency_response_metrics}, under classic GFL control, the grid frequency drops to a nadir of 59.46~Hz with a maximum RoCoF over 2~Hz/s. FFR control can raise the frequency nadir, while the maximum RoCoF remains unimproved, indicating limited frequency support during the inertial response stage. By contrast, the proposed GFM control introduces extra inertia and damping, resulting in a nearly 50\% reduction in the maximum RoCoF and a significant improvement in the frequency nadir.

\begin{table}[!t]
\caption{The rate of change of frequency (RoCoF) and frequency nadir under different control schemes}
\label{tab:frequency_response_metrics}
\centering
\footnotesize
\setlength{\tabcolsep}{2pt}
\renewcommand{\arraystretch}{1.15}
\begin{tabular*}{\columnwidth}{@{\extracolsep{\fill}}cccc@{}}
\hline
\rule{0pt}{2.6ex}
Control scheme & GFL (Classic) & GFL (with FFR) & Proposed\\[2pt]
\hline
\rule{0pt}{2.6ex}
Maximum RoCoF ($\mathrm{Hz/s}$) 
       & $2.054$      & $2.057$       & $1.058$ \\[2pt]
Frequency nadir ($\mathrm{Hz}$)
       & $59.46$       & $59.70$        & $59.86$ \\[3pt]
\hline
\end{tabular*}
\end{table}

\section{Conclusion}
This paper develops an AIDC GFM control scheme for grid support. VSG-based GFM control is implemented in both GSC and LSC to support the external grid and the internal IT loads, respectively.  To exploit the flexibility from both IT loads and UPS ES, 
a coordinated support strategy is developed based on dc-link voltage regulation, with control parameters determined by the power-sharing ratio. To capture the IT-load dynamics in this strategy, an aggregated first-order response model is further developed for multi-server IT loads.

Simulation results conducted on a 100-MW grid-connected AIDC demonstrate the effectiveness of the proposed scheme in coordinating IT-load flexibility and UPS energy storage for grid frequency support. Specifically, the coordinated strategy enables IT loads and UPS ES to share active power according to the prescribed ratio in response to dc-link voltage variations without requiring external power commands or additional communication. Compared with conventional GFL and GFL-FFR controls, the proposed scheme improves the frequency nadir and reduces the maximum RoCoF through enhanced inertial and damping responses.

Future work would investigate the economic trade-off between the IT load performance penalties and the UPS battery degradation to determine the optimal power-sharing ratio.

%\emph{\textcolor{blue}{Future work would investigate the economic trade-off between IT-load flexibility and ES utilization from a system-level perspective, considering workload-related costs, battery degradation, and operational constraints such as QoS requirements and state-of-charge (SOC) limits.}}

\bibliographystyle{IEEEtran}
\bibliography{reference}

\appendices
\numberwithin{equation}{section}
\renewcommand{\theequation}{\Alph{section}-\arabic{equation}}

\ifCLASSOPTIONcaptionsoff
  \newpage
\fi

% \begin{IEEEbiography}[{\includegraphics[width=1in,height=1.25in,clip,keepaspectratio]{Fig/zhiyi_zhao.jpg}}]{Zhiyi Zhao} Zhiyi Zhao received the M.S. degree from Tsinghua University in 2026 and the B.Eng. degree from South China University of Technology in 2024, both in electrical engineering. He is currently working toward the Ph.D. degree at Imperial College London. His research interests include system inertia management.
% \end{IEEEbiography}
% \vspace{-0.9cm}

\end{document}